\documentclass[a4paper]{jpconf}
\usepackage{graphicx}
\begin{document}
\title{Radio Emission in Atmospheric Air Showers: First Measurements with
LOPES-30} 

\author{
P~G~Isar$^{a,1}$, S~Nehls$^{a}$, W~D~Apel$^{a}$, T~Asch$^{b}$, F~Badea$^{a}$,
L~B\"ahren$^{c}$, K~Bekk$^{a}$, A~Bercuci$^{d}$, M~Bertaina$^{e}$,
P~L~Biermann$^{f}$, J~Bl\"umer$^{a,g}$, H~Bozdog$^{a}$, I~M~Brancus$^{d}$,
M~Br\"uggemann$^{h}$, P~Buchholz$^{h}$, S~Buitink$^{i}$, H~Butcher$^{c}$,
A~Chiavassa$^{e}$, F~Cossavella$^{g}$, K~Daumiller$^{a}$, F~Di~Pierro$^{e}$,
P~Doll$^{a}$, R~Engel$^{a}$, H~Falcke$^{c,f,i}$, H~Gemmeke$^{b}$,
P~L~Ghia$^{j}$, R~Glasstetter$^{k}$, C~Grupen$^{h}$, A~Hakenjos$^{g}$,
A~Haungs$^{a}$, D~Heck$^{a}$, J~R~H\"orandel$^{g}$, A~Horneffer$^{i}$,
T~Huege$^{a}$, K~H~Kampert$^{k}$, Y~Kolotaev$^{h}$, O~Kr\"omer$^{b}$,
J~Kuijpers$^{i}$, S~Lafebre$^{i}$, H~J~Mathes$^{a}$, H~J~Mayer$^{a}$,
C~Meurer$^{a}$, J~Milke$^{a}$, B~Mitrica$^{d}$, C~Morello$^{j}$,
G~Navarra$^{e}$, A~Nigl$^{i}$, R~Obenland$^{a}$, J~Oehlschl\"ager$^{a}$,
S~Ostapchenko$^{a}$, S~Over$^{h}$, M~Petcu$^{d}$, J~Petrovic$^{i}$,
T~Pierog$^{a}$, S~Plewnia$^{a}$, H~Rebel$^{a}$, A~Risse$^{l}$, M~Roth$^{a}$,
H~Schieler$^{a}$, O~Sima$^{d}$, K~Singh$^{i}$, M~St\"umpert$^{g}$,
G~Toma$^{d}$, G~C~Trinchero$^{j}$, H~Ulrich$^{a}$, J~van Buren$^{a}$,
W~Walkowiak$^{h}$, A~Weindl$^{a}$, J~Wochele$^{a}$, J~Zabierowski$^{l}$,
J~A~Zensus$^{f}$, D~Zimmermann$^{h}$\\ 
\vspace{2mm}
  LOPES COLLABORATION \\
}

\address{
$^{a}$ Institut\ f\"ur Kernphysik, Forschungszentrum Karlsruhe, Germany\\
$^{b}$ IPE, Forschungszentrum Karlsruhe, Germany\\
$^{c}$ ASTRON Dwingeloo, The Netherlands\\
$^{d}$ NIPNE Bucharest, Romania\\
$^{e}$ Dpt di Fisica Generale dell'Universit{\`a} Torino, Italy\\
$^{f}$ Max-Planck-Institut f\"ur Radioastronomie, Bonn, Germany\\
$^{g}$ Institut f\"ur Experimentelle Kernphysik, Uni Karlsruhe, Germany,\\
$^{h}$ Fachbereich Physik, Universit\"at Siegen, Germany \\
$^{i}$ Dpt of Astrophysics, Radboud Uni Nijmegen, The Netherlands\\
$^{j}$ Ist di Fisica dello Spazio Interplanetario INAF, Torino, Italy\\
$^{k}$ Fachbereich Physik, Uni Wuppertal, Germany \\
$^{l}$ Soltan Institute for Nuclear Studies, Lodz, Poland \\
$^{1}$ on leave of absence from Institute of Space Sciences, Bucharest,
Romania \\ 
}  

\ead{Gina.Isar@ik.fzk.de} 

\begin{abstract}
When Ultra High Energy Cosmic Rays (UHECR) interact with particles in the
Earth's atmosphere, they produce a shower of secondary particles propagating
toward the ground. LOPES-30 is an absolutely calibrated array of 30
dipole antennas investigating the radio emission from these showers in detail
and clarifying if the technique is useful for large-scale
applications. LOPES-30 is co-located and measures in coincidence with the air
shower experiment KASCADE-Grande. Status of LOPES-30 and first measurements
are presented.   
\end{abstract}

\section{Introduction}
The LOPES (LOFAR Prototype Station) experiment co-located with the 
KASCADE-Grande experiment (an extended set-up of the KArlsruhe Shower 
Core and Array DEtector - KASCADE) at Forschungszentrum Karlsruhe, Germany, 
measures the radio emission of air showers in the 40 -- 80\,MHz frequency
range. KASCADE-Grande provides the trigger information and well-calibrated
parameters of the air shower properties in the energy range from a few PeV to 1
EeV. LOPES-30 with a maximum baseline of approximately 260\,m is the extension
 of the initially installed 10 LOPES antennas (LOPES-10)
\cite{Falcke2005,Haungs2006} by the addition of 20 antennas. The antennas now
have absolute calibration, and the array provides a larger sampling area to
the radio signal of a single event compared with the original LOPES-10
set-up. This provides the possibility for a more detailed investigation of the
radio signal on a single air shower basis, in particular of its lateral
extension.   
 
\section{LOPES-30: First detection of radio pulses from air showers}
For the present measurements, all antennas are equipped with
dipoles in East-West direction, measuring the single polarization of
the radio emission. The radio signal
is filtered by a band-pass filter and afterwards digitised with 12-bit ADCs,
working with 80\,MHz, allowing 2nd Nyquist sampling of the signal. From
KASCADE-Grande, LOPES-30 receives a trigger corresponding to
approximately 10\,PeV primary energy. Under this trigger condition, 819\,$\mu$s
of data from the memory buffer of all antennas are stored on a central
DAQ-PC. The main aim in processing the measured LOPES radio raw
signals is to reconstruct the radio field strength of the pulse emitted by an
individual air shower event. The final observable is the value of the
so-called Cross-Correlation beam or CC-beam which combines and averages the
data of a set of antennas to achieve the radio pulse height. For a detailed
description of the hardware and of the necessary steps for the data
processing, see \cite{Horneffer2006,Horneffer2005}. Selecting good radio
events we require a coherent pulse, an expected position of the pulse in time,
and an approximately similar pulse shape in all antennas.   

Since summer 2005, LOPES-30 is acquiring data, measuring the radio
emission in atmospheric air showers and is looking for correlations with
shower parameters like arrival direction, primary particle energy, and
mass. Each single antenna has an absolute calibration using a commercial
reference antenna \cite{Nehls2005}. The calibration leads to the
frequency-dependent amplification factors (see Figure 1, left), representing
the amplification of the electronics. These correction factors are applied to
the measured signal strength resulting in the true electric field strength
which can be compared to the values predicted by simulations. By these
calibration procedures variations between different antennas are corrected
for. Remaining differences (due to variations in the individual antenna gains)
are estimated to be less than 25\,$\%$, which provides also a first estimate
of a systematic uncertainty in the measurements.  

As an example (see Figure 2) we display an event detected by LOPES-30 in
December 2005 of a primary energy of $E_0 \approx 1.6 \cdot 10^{17}$ eV, a
geomagnetic angle (the angle between the shower axis and the Earth magnetic
field) of $36.5^\circ$, and a zenith angle of $15^\circ$. The event shows
clearly the capabilities of LOPES-30: With 30 antennas a very clear and
coherent pulse signal can be detected, separated from the radio frequency
interference (RFI) noise, and reconstructed. Even more, the whole antenna
set-up can be divided in individual clusters of several antennas, and for each
of these antennas the signal can be reconstructed. This procedure allows a
detailed investigation of the radio emission on the basis of individual
events.  
 
\begin{figure}[h]
\centering
\begin{minipage}{6cm}
\includegraphics[width=6.8cm]{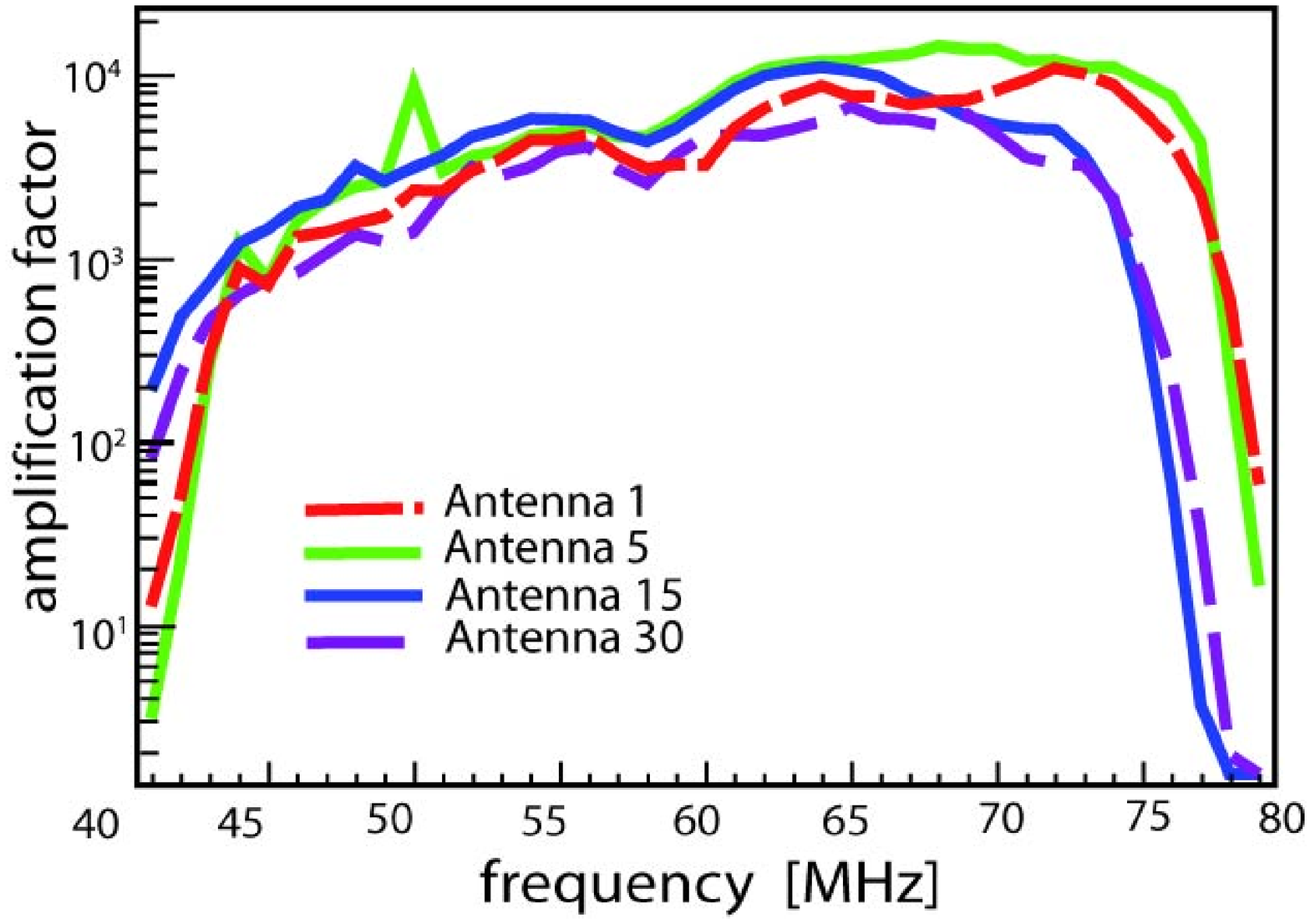}
\end{minipage}\hspace{2cm}%
\begin{minipage}{7cm}
\includegraphics[width=7cm]{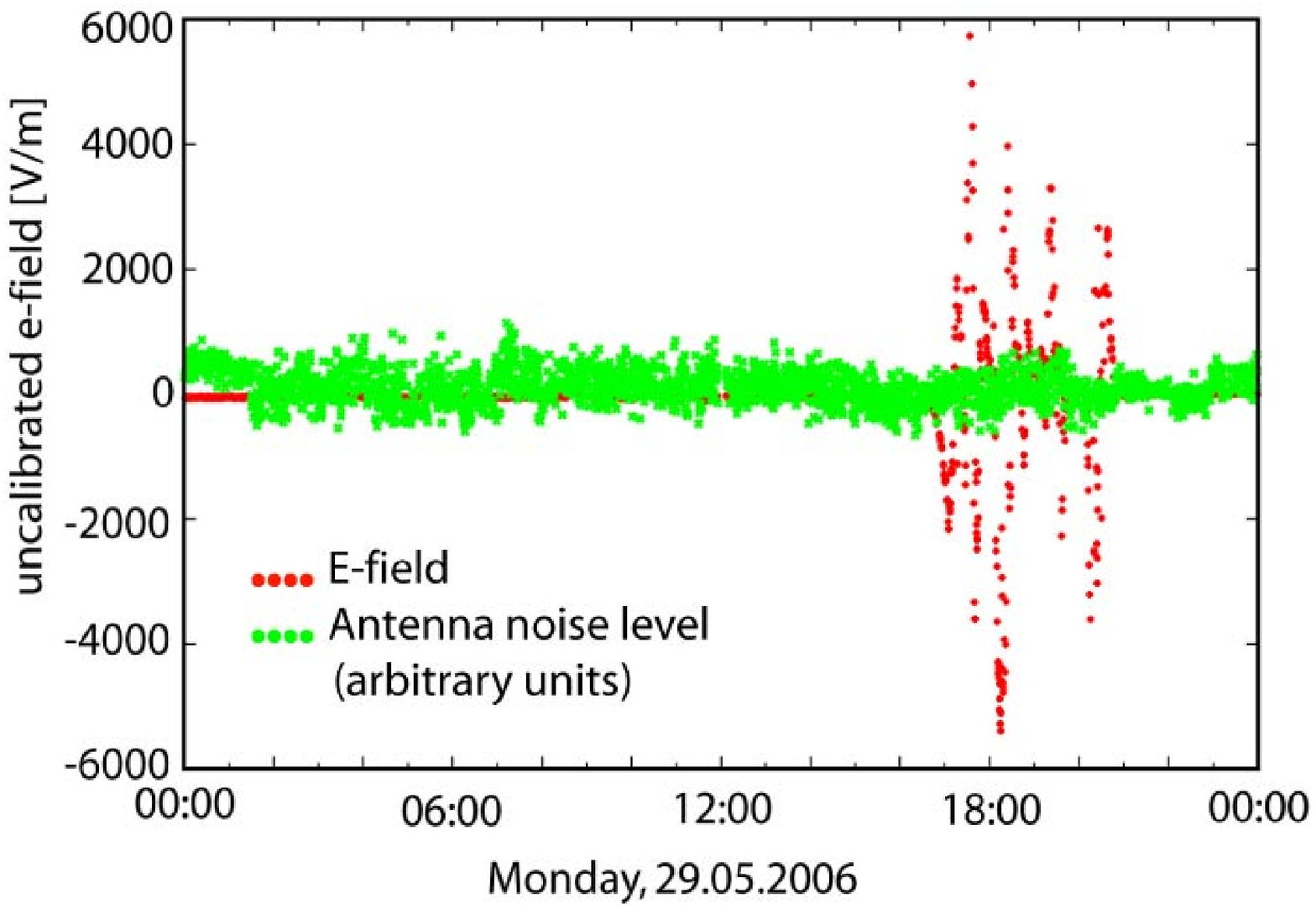}
\end{minipage}\hspace{2cm}%  
\caption{\label{Fig.1} Left: Example for the frequency dependent
amplification factors, including full influence of both, individual antenna
electronics and environmental conditions. The peak at 50\,MHz of antenna 5 is
a faulty measurement due to amateur radio activities. Right: Monitoring
E-field mill data during a day with a strong thunderstorm at Forschungszentrum
Karlsruhe. Shown are the E-field and the relative noise
level of the antennas (arbitrary units): no correlation is found. As shown,
the atmospheric electric field changes dramatically from fair weather
conditions ($\approx$ -160\,V/m) to thunderstorm (ca 18h).}        
\end{figure}

\begin{figure}[h]
\begin{minipage}{2in}
\includegraphics[width=2in]{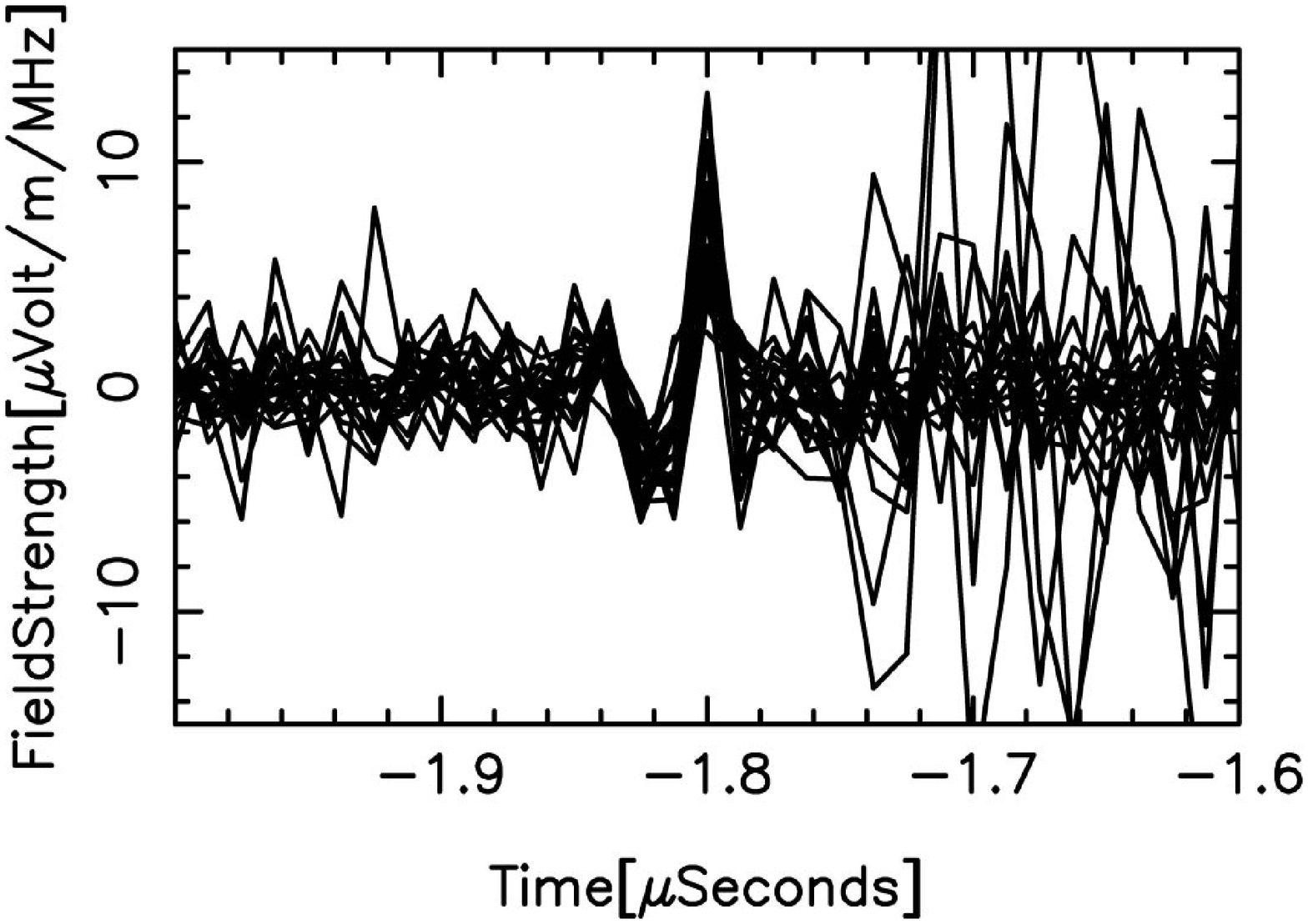}
\includegraphics[width=2in]{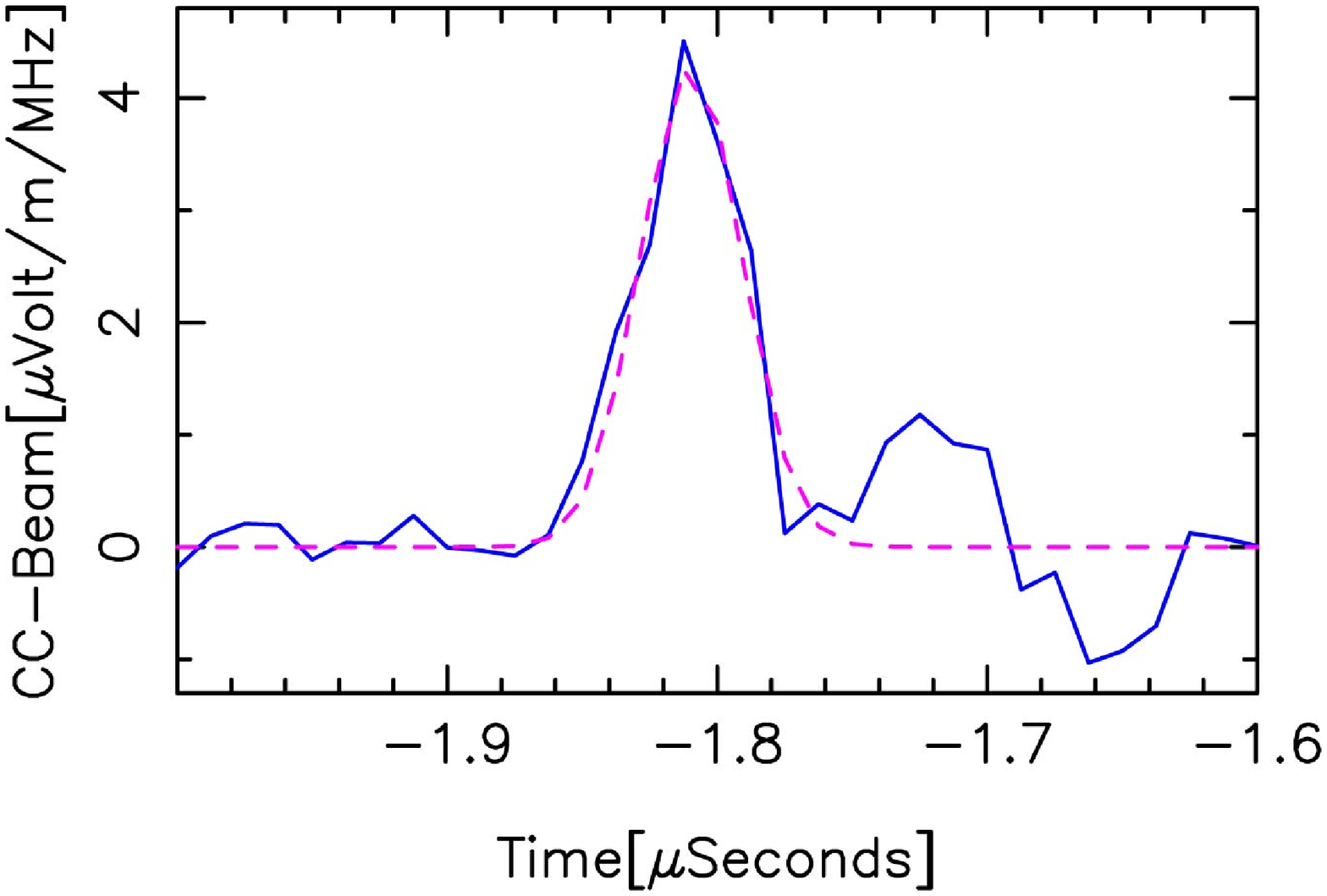}
\end{minipage}\hspace{.2in}%
\begin{minipage}{2in}
\includegraphics[width=2in]{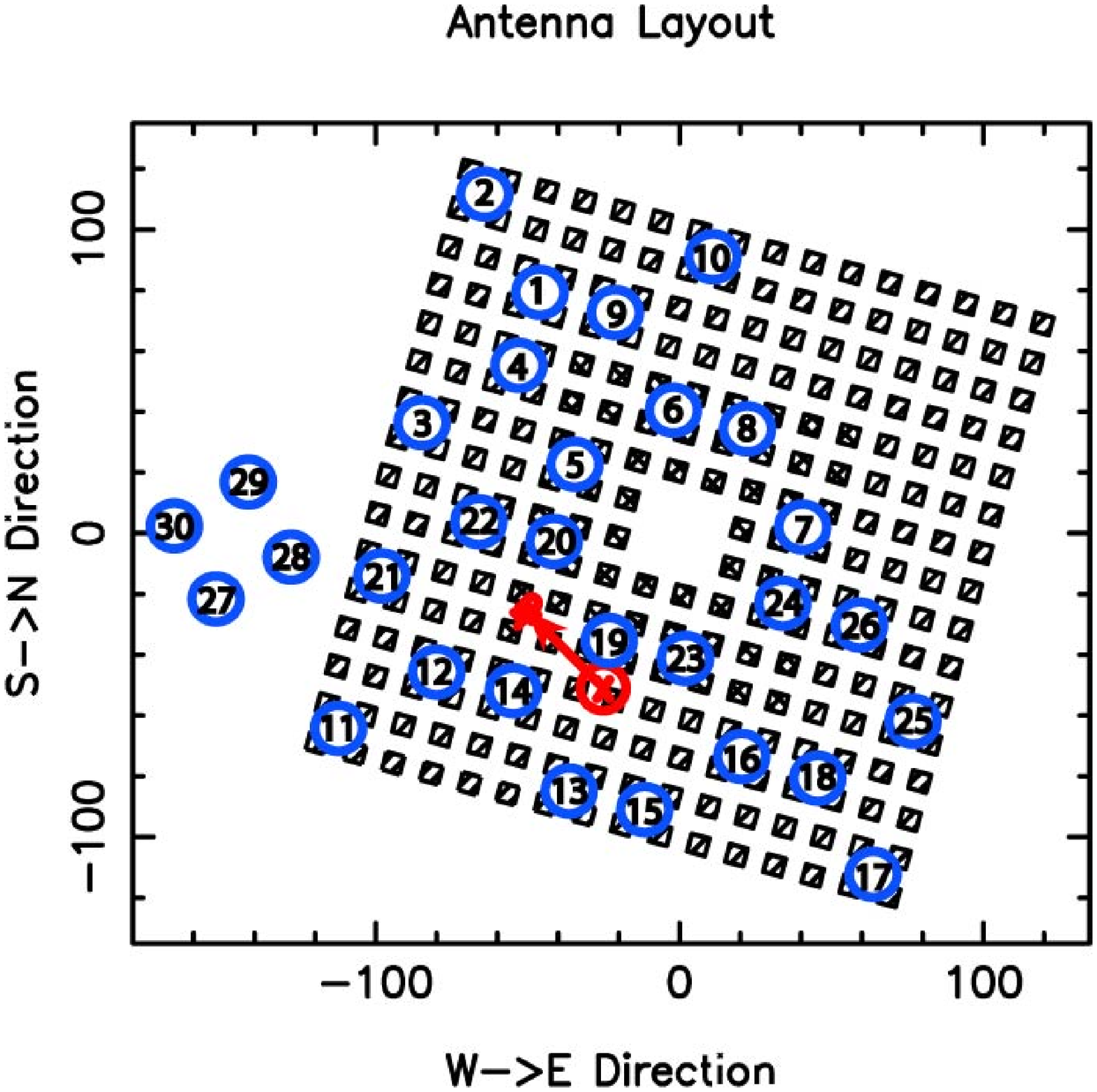}
\end{minipage}\hspace{.2in}%
\begin{minipage}{2in}
\includegraphics[width=2in]{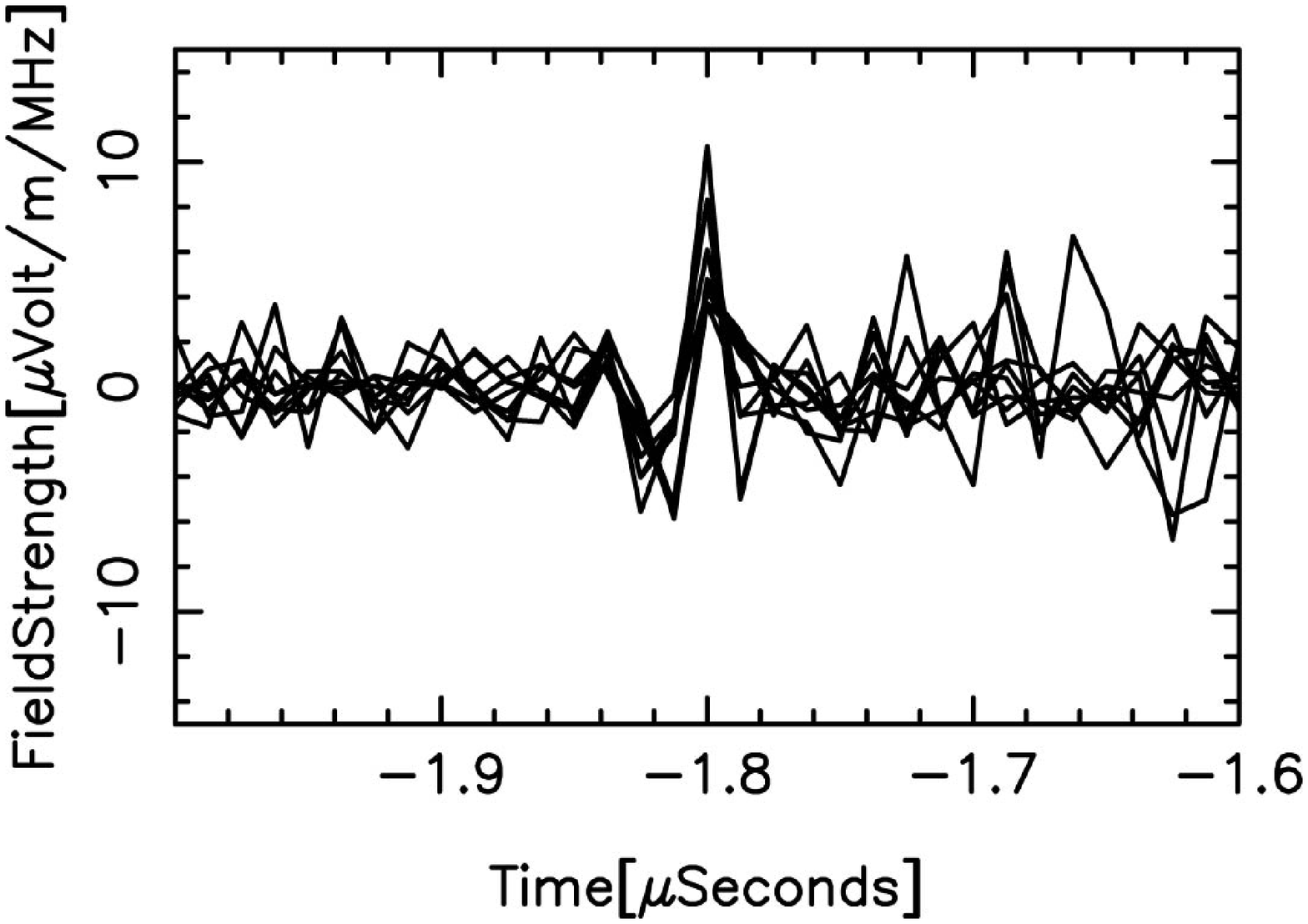}
\includegraphics[width=2in]{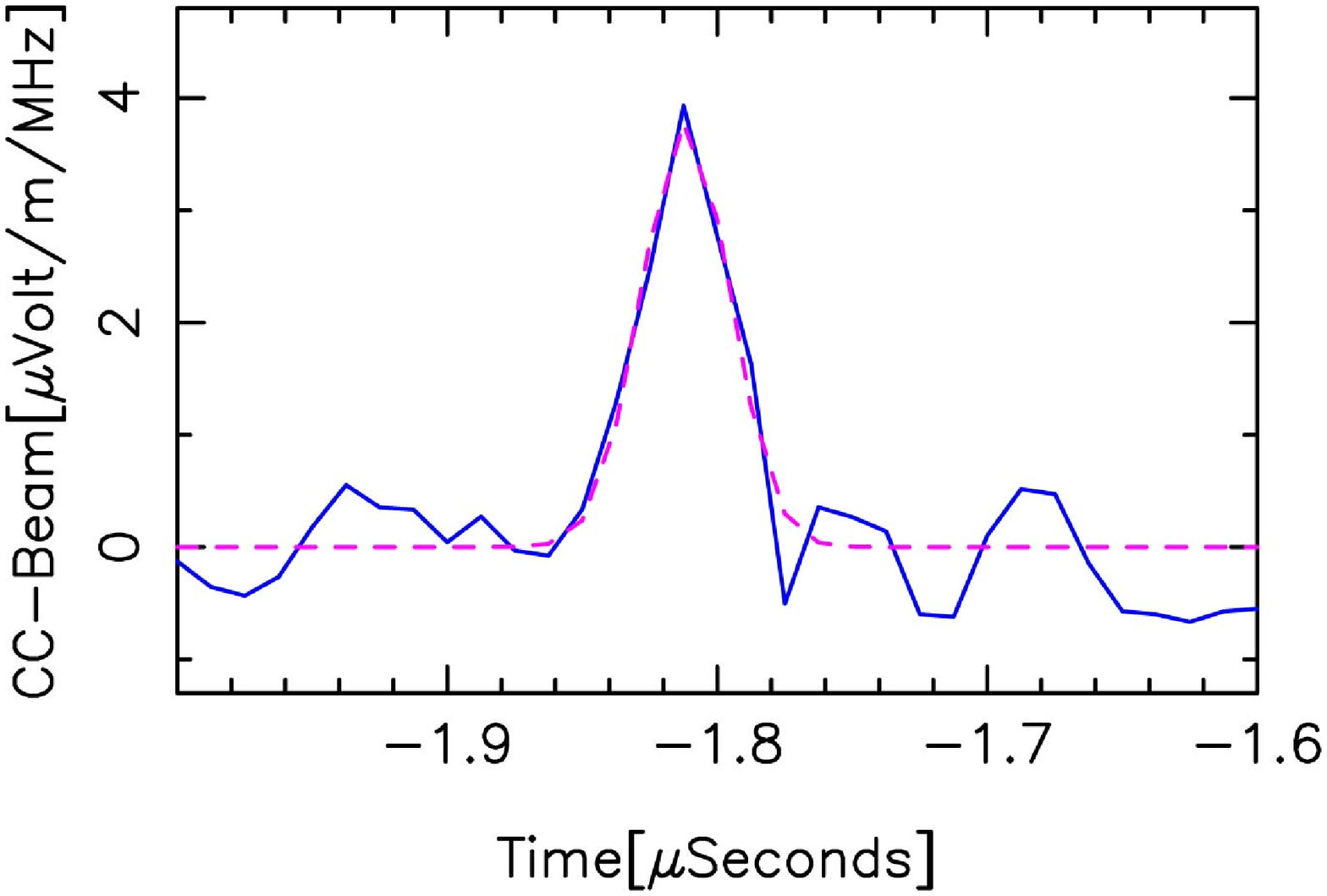}
\end{minipage}\hspace{.2in}%
\caption{\label{Fig.2} Example of an event registered by LOPES-30. Left: Field
Strength of all individual 30 antennas and the result of the Cross
Correlation(CC)-beam forming (Full line: CC-beam. Dotted line: Gaussian
fit). One can clearly distinguish between the coherent radio pulse at
-1.8\,$\mu$s and the detection of incoherent RFI from KASCADE at
-1.7\,$\mu$s. Middle: Antenna layout at the KASCADE array. The arrow indicates
the direction of the incoming cosmic ray shower of this event. Right: The same
event reconstructed by using only a selection of antennas (antenna numbers
from 1 to 10, see layout) and the resulting signal. For antennas positioned
further from the shower core there is a less noise coming from KASCADE
(detector stations), and also the resulting CC-beam value is 12.5\,$\%$ weaker 
than the value obtained by all antennas.} 
\end{figure}

\section{LOPES-30: Environmental Monitoring by an atmospheric E-field mill}
During the LOPES-30 measurements, we put emphasis on monitoring environmental
conditions by measuring the static electric field and by recording parameters
of nearby weather stations. Atmospheric conditions, in particular E-field
variations during thunderstorms, might distort the radio emission during the
shower development, and the measurement of the radio pulses
\cite{Buitink2006}. By monitoring the environmental conditions, and comparing
them with the antenna noise level as well as with the detected air
shower radio signals, correlations can be investigated and corrected for. 

The atmospheric electric field is monitored by a so-called field mill. It
utilises a reciprocating shutter electrically connected to the ground, placed
between the external field and the stationary metal sensor electrodes. This
results in a low-frequency voltage proportional to the low-frequency ($\le$
10\,Hz) electric field. These atmospheric electric field measurements are used
for assessing the local rainfall, lightning hazard and thunderstorms research
(see Figure 1, right). The data of the field mill get continuously stored at
intervals of 1 second in a database, and are investigated for correlations
with the LOPES data measurements. As depicted in the right panel of Figure 1,
between the E-field and the relative noise level of the antennas (arbitrary
units): there is no correlation found.      

\section{Application of radio air shower detection in large UHECR experiments}
One of the main goals of the LOPES project is to pave the way for an
application of this ``re-discovered'' air shower detection technique to large
UHECR experiments like LOFAR (LOw Frequency ARray) and the Pierre Auger
Observatory. In parallel to the measurements at KASCADE-Grande with LOPES-30,
we follow this aim by optimizing the antenna design for such an
application. Additionally, the optimum frequency range, depending on the local
noise, and an adequate filtering is investigated. All these efforts are part
of LOPES$^{\rm STAR}$~\cite{Gemmeke2005}, a ``Self Triggered Array of Radio
detectors for LOPES'', where the measuremets will be compared with
LOPES-30. The possibilities of a self-triggering antenna system and an online
beam forming are also studied setting up a test array at the South part of the
Pierre Auger Observatory.    
 
\section{Outlook}
LOPES continuously takes data in coincidence with KASCADE-Grande. Recently, a
number of antennas have been reconfigured for measurements of the North-South
polarization direction. Measuring both polarizations of the radio emission
will allow to verify the geosynchrotron effect, as the dominant emission
process in atmospheric air showers.  

Recent theoretical and simulation studies \cite{Huege2006,HuegeFalcke2003},
using a sophisticated Monte-Carlo technique, together with the absolute
calibration of the antennas, allow us to compare predictions and radio
emission measured by LOPES-30. 

Monitoring the atmospheric E-field and further environmental
conditions allow to investigate the influence of thunderstorms to the
measurements.   
  
The radio technique can be applied to existing cosmic ray experiments as
well as to large digital radio telescopes like LOFAR and SKA (Square Kilometer
Array), providing a large detection area for UHECR. First approaches using the
radio technique at the Pierre Auger Observatory and at a first LOFAR station
are in progress. Mandatory experience for such applications will be provided
by the LOPES-30 measurements.

\end{document}